\documentclass[a4paper,12pt]{article}
\usepackage[final]{graphics}
\usepackage{epsfig}
\usepackage{amsmath, amssymb}
\usepackage{cite}
\usepackage{epsfig}

\newcommand{\bo}{\raise-1mm\hbox{\Large$\box$}}
 \numberwithin{equation}{section}
 \begin{document}
 \title{Kaluza-Klein reduction for the Unruh brane}
 \author{
 David Jennings \footnote{D.Jennings@damtp.cam.ac.uk}, \\
 \normalsize \em Department of Applied Mathematics and Theoretical Physics,\\
 \normalsize \em CMS, University of Cambridge, Cambridge, CB3 0WA, UK. \\
 }
 \maketitle
 \abstract{We present a Kaluza-Klein reduction for the Unruh response that a brane, in an AdS bulk would experience due to its acceleration. The Unruh radiation is realised as particle production on the brane for each of the Kaluza-Klein modes. A correction, coming from the massive modes, is calculated for the standard Unruh effect in Minkowski spacetime and it is found that the response coming from these modes is exponentially suppressed.}
 \newpage
 \section{Introduction}
Recently much attention has been given to brane models of our universe. These models draw on ideas from string theory and provide novel settings in which our four-dimensional universe is embedded in a higher dimensional spacetime\cite{randall }. The standard model particles are constrained to the brane, while gravity is free to probe the full spacetime. The usual Newtonian gravity arises from the zero mode for gravity being localized on the brane. In addition to this zero mode there are also massive Kaluza-Klein modes that provide correction terms at small scales, and open up the possibility for experimental testing.\\
Much work has been done on quantum fields in such brane models \cite{saharian, naylor, kob, setare1,setare2} and the potential effects that they can have on cosmology, where the signature of the extra dimensions is carried in the Kaluza-Klein modes that exist as massive fields on the brane.\\
It has been argued \cite{paper3, zhang} that a brane universe should perceive an Unruh effect in which the bulk vacuum state is a thermal one, due to the brane following an accelerated trajectory through its bulk spacetime. In the case of an Anti de Sitter (AdS) bulk, the Unruh effect only appears above a critical acceleration in the spacetime. The absence of radiation below this critical acceleration may be understood if the accelerated trajectory in AdS is embedded in a flat spacetime of a higher dimension. In this case the sub-critical trajectories are not hyperbolic trajectories in the flat spacetime.\\
It has been shown \cite{deser1} that black hole radiation, and its associated thermal properties, can be realised by considering global embedding Minkowskian spacetimes (GEMS), in which the thermal properties of the black hole arise as the result of the observer in the curved spacetime being mapped onto a hyperbolic trajectory in the GEMS. On such a trajectory, the observer experiences Unruh radiation with a temperature proportional to their acceleration in the embedding spacetime. This mapping of particle production in a curved spacetime into Unruh radiation has also been demonstrated for de Sitter spacetime \cite{deser2}.
In this paper we shall consider the reverse proceedure. We shall consider the response of a brane universe on a constant acceleration trajectory in a bulk AdS spacetime, and analyse the excitations in terms of particle production for Kaluza-Klein modes in the lower dimensional brane. We organise this paper as follows: In section 2 we consider a two brane model with an AdS bulk written as a warped product, perform a mode decomposition and analyse the solution and mass spectrum for the case of a conformally coupled, massless scalar field. In section 3 we quantize the mode solutions on the brane and discuss the detector response function for a brane following the three basic accelerated trajectories in AdS spacetime. In the case of a Minkowski brane we compute a correction term, coming from the massive modes, for the standard Unruh effect. Section 4 contains the conclusions and discussion of further work.
\section{Scalar field in the bulk}
We shall consider the two brane model with an $(N+1)$-dimensional bulk Anti de Sitter spacetime. The metric in this case may be written as a warped product
\begin{eqnarray}\label{metric}
 ds^2 &=& A(y)^2 ds^2_{\mbox{\tiny brane}} - dy^2
\end{eqnarray}
with one brane situated at $y=0$ and the other at $y=L$, and we impose $Z_2$-symmetry across the branes.\\
Any observer moving along a trajectory $y=y_0=$ constant in this spacetime will have a constant proper acceleration
\begin{eqnarray}
 a^2&=& \left ( \frac{A'}{A} \right )^2(y_0)
\end{eqnarray}
where the dash denotes differentiation with respect to $y$. Consequently, if the brane is not located at a maximum or minimum of the warping function $A(y)$, then it will have an acceleration in the bulk spacetime.\\
The metric of the brane is assumed to be homogeneous and isotropic, and in addition there is an effective cosmological constant $\Lambda_b$ induced on the brane, which depends on the bulk cosmological constant $\Lambda $ and the brane tension $ \sigma  $, through the equation
\begin{eqnarray}
 \Lambda_b &=& \frac{ \kappa ^4 \sigma^2}{36} + \frac{\Lambda }{6}.
\end{eqnarray} 
The form that the metric function $A(y)$ takes will depend on whether $\Lambda _b$ is positive, negative or zero.\\
We consider a bulk scalar field, of mass $M$, and obeying the field equation
\begin{eqnarray}
(\Box + M^2 + \zeta R +c_1 \delta (y) + c_2 \delta (y-L)) \phi (X) &=& 0 
\end{eqnarray}
where $ \zeta $ is the coupling to gravity, $R$ is the Ricci scalar for the AdS bulk and $c_1,c_2$ are mass parameters \cite{saharian2} for the two branes. In what follows we analyse the decomposition of this field.
\subsection{Mode Decomposition}
For the warped product (\ref{metric}), we may decompose the bulk d'Alembertian into
\begin{eqnarray}
 \Box &=& \frac{ \Box _b}{A^2} - \frac{1}{A^N}\partial_y (A^N \partial_y)
\end{eqnarray}
where $\Box _b$ is the d'Alembertian on the brane. We now consider the mode decomposition $\phi (X) = \sum_n \phi _n(x^ \mu ) f_n (y)$, which gives rise to the equations
\begin{eqnarray}
(\Box _b + m_n^2)\phi _n (x ^\mu ) &=&0 
\end{eqnarray}
and 
\begin{eqnarray}
 f_n '' + \frac{N A'}{A} f_n - ( \zeta R + M^2 -\frac{m_n^2}{A^2} -c_1 \delta (y) - c_2 \delta (y-L)) f_n = 0.
\end{eqnarray}
The Ricci scalar for the bulk spacetime is composed of two parts. The first part comes from the regular AdS geometry, and the second is a delta function contribution coming from the branes.\\
It is natural to change to coordinates for which the metric is conformally simple.
By changing variables, we may write the metric in the form
\begin{eqnarray}
 ds^2&=& A^2 (ds^2_{\mbox{\tiny brane}} - dz^2) 
\end{eqnarray}
where
\begin{eqnarray}
 z&=&\int \frac{dy}{A(y)}
\end{eqnarray}
and so we may express the bulk Ricci scalar as
\begin{eqnarray}\label{ricci}
R &=& \frac{1}{A^2}  ( R_b - N( \frac{2 \ddot{A}}{A} +(N-3)\left ( \frac{\dot{A}}{A} \right )^2 ) )
\end{eqnarray}
where $R_b$ is the Ricci scalar for the brane, and a dot denotes differentiation with respect to the variable $z$. The term $\ddot{A}$ gives rise to delta function contributions once we impose $Z_2$-symmetry across the branes. We find that the Ricci scalar is
\begin{eqnarray}
 R&=& -N(N+1)k^2 - 4N\frac{A'}{A} (\delta (y) - \delta (y-L)) 
\end{eqnarray}
where $k$ is the curvature of AdS bulk.\\
The transverse mode equation then becomes
\begin{eqnarray}
 f_n '' + \frac{N A'}{A} f_n + ( \zeta N(N+1)k^2  - M^2 +\frac{m_n^2}{A^2} +&& \nonumber \\
(c_1-\frac{4 \zeta N A'}{A} ) \delta (y) + (c_2 +\frac{4 \zeta N A'}{A} )\delta (y-L)) f_n &=& 0.
\end{eqnarray}
The boundary conditions imposed on $f_n$ are obtained by integrating this equation over the brane locations, and we may consider both twisted solutions, $f_n(-y)=-f_n(y)$, and untwisted solutions, $f_n(-y)=f_n(y)$. In the twisted case the boundary conditions are Dirichlet.\\
For general scalar mass and coupling to gravity this equation is normally solvable in terms of hypergeometric functions, but in the case of a conformally coupled, massless scalar field we may solve in terms of elementary functions. In this particular case the equation for $f_n$ may be written as
\begin{eqnarray}
\ddot{f}_n + \dot{f}_n (N-1) \frac{\dot{A}}{A} + \left ( \left (\frac{N^2-1}{4} \right )k^2A^2 + m_n^2 \right )f_n &=& 0 .
\end{eqnarray}
We introduce the function $g_n(z) = A(z)^\beta f_n(z)$, and find, for the choice of $ \beta = \frac{N-1}{2}$ and using equation (\ref{ricci}), that $g_n(z)$ is determined from
\begin{eqnarray}
 \ddot{g}_n + ( m_n^2-\zeta_c R_b )g_n(z) &=& 0
\end{eqnarray}
where $\zeta _c$ is the conformal coupling to gravity. Since we assume that the brane has constant curvature the mode solution then takes the form
\begin{eqnarray}
f_n(y) &=& A(y)^{\frac{N-1}{2}} ( B_1 \sin \omega_n z(y) + B_2 \cos \omega _n z(y))
\end{eqnarray}
with $\omega _n^2 = m_n^2 - \zeta _c R_b$. The constants $B_1,B_2$ and the mass spectrum are determined from the boundary conditions, together with the normalisation condition
\begin{eqnarray}
 \int_{-L}^{L} dy A^{N-2}(y)f_n(y) f_{m}(y)&=& \delta _{nm}.
\end{eqnarray}
In the untwisted field case the mass spectrum may be solved explicitly only in the case where $A(0)c_1 = -A(L)c_2$, and the masses are given by
\begin{eqnarray}
 \omega_n &=&  \frac{n \pi }{|z_L - z_0|}
\end{eqnarray}
where $n=0,1,2 \cdots$ and $z_0 = z(y=0), z_L = z(y=L)$. The twisted field solution has exactly the same spectrum, without any restriction on the mass parameters $c_1, c_2$. The normalised eigenfunctions are then
\begin{eqnarray}\label{twisted}
 f_n(z)&=& \frac{A^{(\frac{N-1}{2})}}{\sqrt{|z_L -z_0|}} \sin \omega _n (z - z_0)
\end{eqnarray}
in the twisted case, and
\begin{eqnarray}\label{untwisted}
 f_n(z)&=& \frac{A^{(\frac{N-1}{2})}}{\sqrt{|z_L -z_0|} \sqrt{1 + \alpha_n^2} }(\cos \omega _n (z-z_0) + \alpha _n \sin \omega _n (z - z_0) )\nonumber \\
\alpha _n &=& \frac{A(z_0)c_1}{2 \omega _n}
\end{eqnarray}
in the untwisted case. We now proceed to analyse the response of a brane detector, coupled to the field $\phi (X)$, in terms of Kaluza-Klein modes.
\section{Response Function for the brane}
The response of a detector moving along a general trajectory $X(\tau )$ and coupled to a field $\phi $ is determined by the response function \cite{davies}
\begin{eqnarray}
 F(E)&=& \int_{- \infty}^{\infty} d \tau d \tau ' e^{-i(\tau - \tau ') E} G^+ (X(\tau ), X ( \tau '))
\end{eqnarray}
where $E>0$ and $G^+(X, X')=<0|\phi(X) \phi (X') |0>$ is the Wightman function for the scalar field $\phi $. In analysing the response function for the brane, we proceed by quantizing each $\phi_n(x^ \mu )$ as an $N$ dimensional scalar field \cite{naylor}, and in doing so we may write
\begin{eqnarray}
 G^+(X, X')&=& \sum_n f_n(y)f_n(y')G^+_b (x^ \mu , x '^ \mu ; \omega _n)
\end{eqnarray}
where $f_n$ is given by (\ref{twisted}) or (\ref{untwisted}) and $G^+_b(x^ \mu , x '^ \mu ; m)$ is the Wightman function for a scalar field, of mass $m$, in a spacetime with the metric $ds^2_{\mbox{\tiny brane}}$.\\
An detector constrained to the brane will have a trajectory $X( \tau ) = (x^ \mu ( \tau ), y ( \tau )) = (x^ \mu (\tau ), 0)$. The response function for such a detector will vanish in the twisted field case, and in the untwisted case it is given by
\begin{eqnarray}
 F(E)&=& \sum_n \left ( \frac{A(z_0)^{N-1}}{(|z_L-z_0|)(1+ \alpha ^2_n)} \right ) F_n (E)
\end{eqnarray}
where $F_n(E)$ is the $N$-dimensional response function for a detector on a trajectory $x^\mu (\tau )$ in a spacetime with the metric $ds_{\mbox{\tiny brane}}^2$, and coupled to a scalar field of mass $ \frac{n \pi}{|z_L-z_0|}$. The spectral weight of the mode $n$ is given by $f_n(z_0)^2$.
We may now consider what happens for a brane with a constant proper acceleration. In AdS spacetime there are three distinct classes of accelerated trajectories\cite{3class}, depending on whether the acceleration is greater than, less than or equal to the AdS scale $k$. These trajectories are called super-critical, sub-critical or critical, respectively. In \cite{paper3} it was argued that a brane with constant acceleration would only experience an Unruh effect for super-critical trajectories, which we may now analyse in terms of the lower dimensional picture.\\
A brane moving along a super-critical trajectory has the warping function
\begin{eqnarray}
A(y) &=& \frac{\sqrt{\Lambda _b}}{k} \sinh k (y_h - |y|)
\end{eqnarray}
where $y_h$ is the position of the horizon, and brane metric is that of de Sitter spacetime. The proper acceleration of the brane is seen to be
\begin{eqnarray}
 a &=& k\coth ky_h > k 
\end{eqnarray}
and the response per unit time, of a comoving detector on the brane is then
\begin{eqnarray}
\frac{dF(E)}{d \tau } &=& \sum_n \left ( \frac{A(z_0)^{N-1}}{(|z_L-z_0|)(1+ \alpha ^2_n)} \right ) \int_{-\infty}^\infty d \Delta \tau G^+_{\mbox{\tiny dS}, n}(\Delta \tau )e^{-i \Delta \tau  E } \nonumber
\end{eqnarray}
where $G^+_{\mbox{\tiny dS}, n}(\Delta \tau )$ is the de Sitter Wightman function for the n$^{\mbox{\tiny th}}$ mode evaluated along a comoving trajectory. Each mode will separately undergo particle creation in the de Sitter spacetime, and for the massive modes we obtain deviations \cite{ds1, ds2} from the standard thermal spectrum spectrum with temperature $\frac{\sqrt{ \Lambda _b }}{2 \pi }$.\\
For a brane moving along a sub-critical trajectory the warp factor is
\begin{eqnarray}
 A(y)&=& \frac{ \sqrt{| \Lambda _b|}}{k} \cosh k (c - |y|) ,
\end{eqnarray}
its proper acceleration is $k\tanh kc <k$ where $c$ is a constant, and the brane geometry is AdS spacetime. A comoving detector on the brane will not register excitations from any of the modes.\\
In the critical acceleration case $A(y) = e^{-ky}$, and the brane, located at $y=0$, has acceleration $a=k$ in the bulk spacetime. We take the geometry of the brane to be Minkowski spacetime and, as in the sub-critical case, there are no excitations for a comoving detector.
\subsection{Correction to the Unruh effect on a Minkowski brane}
We now wish to consider the contributions that massive modes provide for the Unruh radiation experienced by a detector that accelerates on a Minkowski brane.
The Wightman function for an N-dimensional scalar field with mass $m$ in Minkowski spacetime is given by
\begin{eqnarray}
 G^+(x^\mu , x'^\mu ; m) &=& \frac{1}{2 \pi }\left ( \frac{m}{2 \pi w } \right )^{\frac{N-2}{2}} K_{\frac{N-2}{2}} (m w) \nonumber \\
w^2 &=& (\mathbf{x} - \mathbf{x'})^2 - (t -t' - i \epsilon )^2
\end{eqnarray}
where $K_\nu (z)$ is the modified Bessel function, where we have implemented the standard $i \epsilon $ regularization.\\
In the case of large Kaluza-Klein masses we may perform an asymptotic expansion for the Wightman function and calculate explicitly the contribution to the response function from these massive modes in the case of a detector moving with constant acceleration in the brane.
The response rate function for a detector on a stationary trajectory in the brane is given by
\begin{eqnarray}
\frac{dF(E)}{d \tau } &=& \sum_n f_n(z_0)^2 \frac{dF^M_n}{d \tau }
\end{eqnarray}
where $\frac{dF^M_n}{d \tau }$ is the response rate function in the $N$-dimensional Minkowski spacetime for a particle of mass $m_n$. For a detector moving along a hyperbolic trajectory in the brane, the massless mode contribution is given by \cite{takagi}
\begin{eqnarray}
 \frac{dF^M_0}{d \tau }(E)&=& \frac{p_N(E)}{ e^{\frac{2 \pi E}{a}} - (-1)^N} 
\end{eqnarray}
where $p_N(E)$ is a polynomial of degree $N-3$ and $a$ is the $N$-dimensional acceleration of the detector.\\
The response rate from the mode of mass $m_n$ may be expressed as
\begin{eqnarray}
\frac{dF^M_n}{d \tau } &=& \frac{2}{E | \Gamma (\frac{iE}{a})|^2( e^{\frac{2 \pi E}{a}} -1)} \int \frac{d^{N-2}k}{(2 \pi )^{N-2}} K_{\frac{iE}{a}} (s)^2
\end{eqnarray}
where $(as)^2 = m_n^2 + |\mathbf{k}|^2$. If we assume that $m_n$ is large and make use of the asymptotic expansion for the modified Bessel function then we find that 
\begin{eqnarray}
 \frac{d F^M_n}{d \tau }&=& \frac{1}{2 m_n }\left ( \frac{m_na}{4 \pi } \right ) ^{\frac{N-2}{2}}e^{-\left (\frac{ \pi E + 2 m_n}{a} \right ) }.
\end{eqnarray}
Consequently, for $n=1$ we obtain that the first correction to the standard massless response rate in 4-dimensional Minkowski spacetime is given by
\begin{eqnarray}
\frac{dF_1}{d \tau } &=& \frac{a}{ 8 \pi (|z_L- z_0|)(1+ \alpha_1^2)}e^{-\left ( \frac{2 \pi }{a (|z_L-z_0|)} \right ) } e^{-\frac{ \pi E }{a} }.
\end{eqnarray}
These results also serve to illustrate that in general the higher massive modes will be exponentially suppressed in the response of an accelerated detector, as we would expect.
\section{Conclusion}
In this paper we analysed the response function for a brane universe, moving along stationary trajectories in Anti de Sitter spacetime.\\
We performed the mode decomposition for a scalar field in a general warped product spacetime, with two $Z_2$-symmetric branes, where we included mass parameters on the branes that were coupled to the field. For the conformally coupled massless scalar field explicit solutions were obtained for the transverse modes, together with a simple mass spectrum in the case where the brane mass parameters satisfy the condition $c_1A(z_0)=-c_2A(z_L)$.\\
By quantizing the Kaluza-Klein modes on the $N$-dimensional brane we were able to write the response of a brane detector as a sum over the contributions from each mode. In this setting, the absence of an Unruh effect for a detector on a sub-critical or critical trajectory in the bulk spacetime is simply expressed in the lower dimensional setting where the modes do not undergo any particle production. A correction was obtained to the standard Unruh effect for detector accelerating on a Minkowski brane, coming from the massive modes, which are exponentially suppressed.\\
This type of analysis has complications if we try to apply it to a general cosmological setting. In the case of a brane moving with a variable matter density, its bulk trajectory is no longer stationary and its acceleration is variable. Two difficulties arise in attempting to analyse the Unruh response on such a brane. On one hand the mode decomposition cannot be performed as before, since it is not possible to write the metric in terms of a warped product. On the other, there is an issue of the trajectory being non-stationary. Substantial work has been conducted on non-stationary responses in Minkowski spacetime \cite{non-stat1, non-stat2}. Instead of considering the contribution from all points along the trajectory, it is necessary to consider a causal response function that only depends on the past history of the detector. In addition, the standard $i \epsilon $ regularization is insufficient and must be modified to give meaningful results. To analyse the Unruh response for a general cosmological brane in a bulk spacetime, it is necessary to tackle these points.
\section*{Acknowledgements}
This work is funded by St. John's College Cambridge.

 \end{document}